\documentclass[12 pt]{article}
\usepackage[dvips]{epsfig}
\usepackage{multirow}
\newtheorem{theorem}{\sc THEOREM}[section]

\newtheorem{lemma}{\sc Lemma}[section]

\newtheorem{prop}{Proposition}[section]
\def\a{\alpha}
\def\b{\beta}
\def\s{\sigma}
\def\xs{\left({x\over \s}\right)}

\def\F{\left[1-e^{-\xs^\b}\right]}
\def\Pa{\left[{\partial\over \partial \theta}\log h(x;\theta)\right]}
\def\pa{{\partial\over \partial \theta}}
\def\pah{{\partial\over \partial \theta}\log h(x;\theta)}
\def\paf{{\partial\over \partial \theta} f(x;\theta)}
\def\paF{{\partial\over \partial \theta} F(x;\theta)}
\def\sr{\sum_{i=1}^r}
\def\zz{\left({x\over\sigma}\right)^\beta}

\begin{document}
\title{Fisher information matrix %Efficiency analysis
 for three-parameter\linebreak
exponentiated-Weibull distribution under type II censoring}

\author{Lianfen Qian\\Department of
Mathematical Sciences, Florida Atlantic University, \\Boca Raton,
FL 33431, U.S.A. \  Email: lqian@fau.edu}

\date{}
\maketitle

\begin{abstract}
This paper considers the three-parameter exponentiated Weibull family
under type II censoring. It first graphically illustrates the shape property of the hazard function. Then, it proposes a simple algorithm for computing the maximum likelihood estimator and derives the Fisher information matrix. The latter one is represented through a single integral in terms of hazard
function, hence it solves the problem of computation difficulty in
constructing inference for the maximum likelihood estimator. Real data analysis is conducted to illustrate the effect of censoring rate on the maximum likelihood estimation.
\end{abstract}

%\begin{keywords}

%%%%%%%%%%%%%%%%%%%%%%%%%%%%%%%%%%%%%%%%%

\section{Introduction}\label{sec:intr}
In testing the reliability of a component, $n$ identical
components are placed on life-testing. The type II censoring
scheme is to stop the test procedure when one observes the $r$th
failure, $r\le n$. Various models have been proposed for lifetime
distribution. Among those lifetime distributions, Weibull
distribution is the most popular used. Based on Weibull
distribution, various generalizations have been studied (Pham and
Lai, 2007). Among those generalizations, one of the families  is
called exponentiated Weibull distribution (EWD), initially
proposed by Mudholkar and Srivastava (1993). EWD family not only
covers the one-parameter exponential family, exponentiated
exponential family as a sub-family, but also covers the most
popular used two-parameter Weibull family as a special sub-family.
One of the nice features of EWD family is that it allows
non-monotonic hazard functions, such as unimodal shaped and
bathtub shaped, appeared in science, engineering and medical
fields. For more shapes of hazard functions, see Mudholkar and
Srivastava (1993). Mudholkar, Srivastava and Freimer (1995)
reanalyzed the bus-motor-failure rate data using EWD family.
% In their paper it did mention that in order to make inference of the
%estimator, one needs the asymptotic normality for the parameter
%estimator, but both papers (Mudholkar, et al. 1993 and 1995) did
%not obtain the asymptotic results.

Singh, Gupta and Upadhyay (2002, 2005) studied the point
estimators of three-parameters for EWD under complete data and
type II censored using various estimation methods such as maximum
likelihood method, Bayes method and generalized maximum likelihood
method. Numerical comparisons were obtained for the point
estimators.
%Still once again, no asymptotic results were obtained.
Ortega, Cancho and Bolfarine (2006) gave an influence
diagnostics in exponentiated Weibull with censored data.
Ortega, et al. (2006) states that ``it is not possible to compute
the Fisher information matrix...''.

In this paper, we  derive a simple expression for the Fisher
information matrix through a single integral of the hazard
function, hence obtain the asymptotic normality of the maximum
likelihood estimator of the three unknown parameters for EWD under
type II censoring.
%Numerical illustration will be given.

\section{Shape property of the hazard function}

\begin{figure}
\epsfig{file=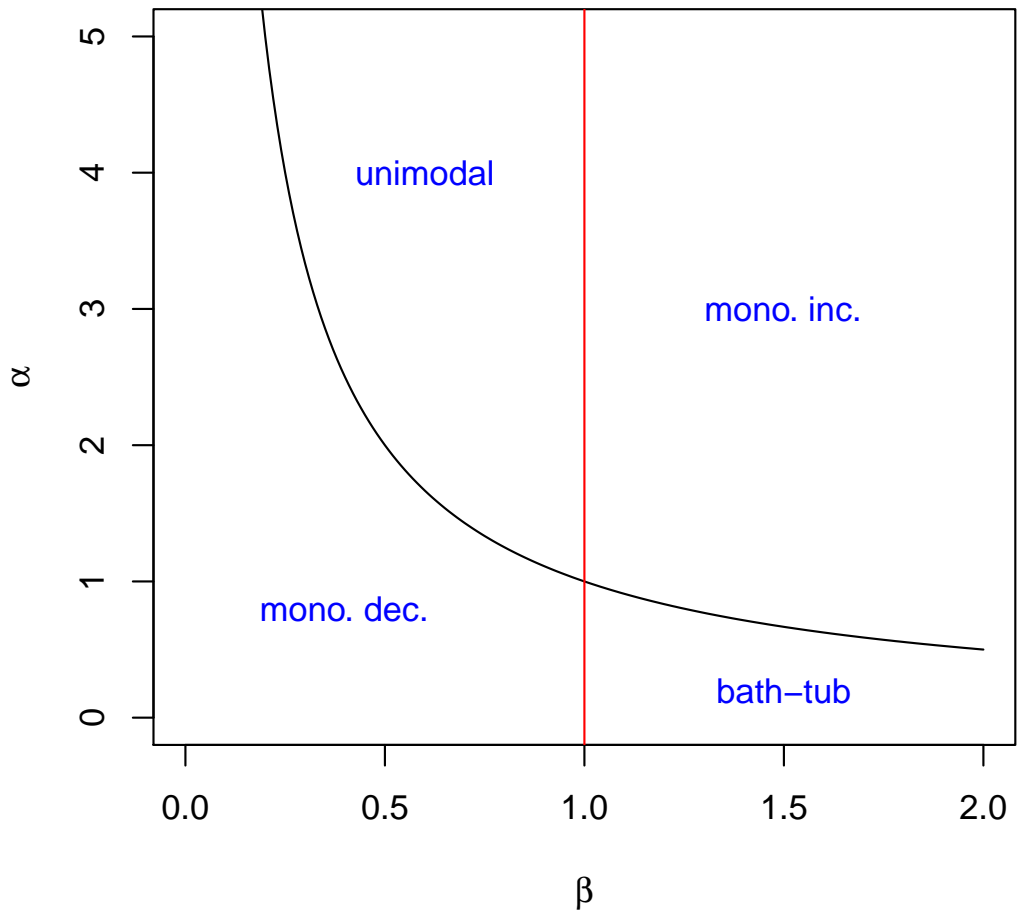, width=.90\linewidth}
\caption{The
graphical display of the four regions which separate the parameter
domain for shape properties of the hazard function, where the curve is for $\alpha\beta=1$. mono. dec=monotonic decreasing, mono. inc.=monotonic increasing.}
\end{figure}

Assume that all $n$ independent components
under testing have cumulative distribution function
$F(x;\theta)$, density function $f(x; \theta)$ and hazard function
$h(x;\theta)$ with parameter vector $\theta\in\Theta\subset
R_{+}^3$, where $\Theta$ is an open set in $R_{+}^3$,
$R_{+}=(0,\infty)$. Let $\theta=(\a,\b,\s)^T$. Then the distribution and density functions
are
$$ F(x;
\theta)=\left[1-e^{-\xs^\b}\right]^\a, \ \ \ x>0
$$
and
$$
f(x;\theta)={\a\b\over \s}\xs^{\b-1} e^{-\xs^\b}\F^\a,  \ \ \
x>0
$$
respectively. Here $\a$ and $\b$ are shape parameters and $\s$ is a scale parameter.
Notice that if $\a=1$, it reduces to the two-parameter Weibull
distribution family. If $\b=1$, it reduces to the exponentiated
exponential family. If $\a=1$ and $\b=1$, it is the one-parameter
exponential family. If $\alpha=1$ and $\beta=2$, it reduces to
Rayleigh distribution and generalized Rayleigh or Burr type X distribution if $\beta=2$.

 The hazard function is the ratio of $f$ and $1-F$. It takes various shapes. To be more precise, the hazard function of EWD is
\begin{equation}\label{eq1}
h(x)={{\a\b\over \s}
\left[1-\exp\left(-\zz\right)\right]^{\alpha-1}\exp\left(-\zz\right)\left({x\over
\sigma}\right)^{\beta-1}\over
1-\left(1-\exp\left(-\zz\right)\right)^\alpha}.
\end{equation}

Mudholkar, Srivastava and Freimer (1995) stated that the monotonicity property of the hazard function is completely determined
by $(\alpha, \beta)$ in the first quadrant. The first quadrant is divided into four regions as shown in Figure 1, where the curves are the boundaries of the four regions. That is $\alpha\beta=1$ and $\beta=1$. It is easy to understand that the shape property of the hazard function is independent of the scale parameter $\s$.  To see this, let $X$ be EWD with parameters $\theta$, denoted by $X\sim EWD(\a,\b,\s)$. Then $X/\s\sim EWD(\a,\b,1)$.
However the shape property with respect to the shape parameters is not easy to see. Mudholkar, Srivastava and Freimer (1995) gave the theorem (their Theorem 2.1), but no detail proof.
We will illustrate the theorem through visualizing the sign of the derivative of
the hazard function. For this purpose, we first  derive
the derivative function of the hazard function, then make comments
on the sign of the derivative function.

\begin{prop} Let
$$
s(z)=\b z \ln
z\left[(z-1)^\a+(\a-z)z^{\a-1}\right]+(\b-1)(z-1)\left[z^\a-(z-1)^\a\right].
$$
Then the sign of the derivative of the hazard function $h$ is the same as the function $s$.
\end{prop}

{\bf Proof:}
Let $z=\exp\left(\zz\right )$. Then $z>1$,
$\ln z=\left({x\over \sigma}\right)^\b $ and $x=\sigma
\left(\ln z\right)^{1\over \b}$. Rewrite $h$ function into a
function in $z$ and denote as $r(z)$. We have, for $z>1$,
$$%\begin{eqnarray*}
 r(z)=h(\sigma \left(\ln
z\right)^{1\over \b}) ={\a\b\left[1-z^{-1}\right]^{\a-1}z^{-1}\left(\ln
z\right)^{\b-1\over \b}\over \s
\left[1-(1-z^{-1})^\a\right]}%\\
=%&=&
{\a\b\over \s}{ (z-1)^{\a-1}\left(\ln
z\right)^{\b-1\over \b}\over
\left[z^\a-(z-1)^\a\right]}.
$$%\end{eqnarray*}
Thus, taking logarithm both sides, we have
$$
\ln r(z)=\ln {\a\b\over \s}+(\a-1)\ln
(z-1)+{\b-1\over \b}\ln\ln z-\ln[z^\a-(z-1)^\a].
$$
Taking derivative, we obtain
\begin{eqnarray*}
{r'(z)\over r(z)}&=&{\a-1\over z-1}+{\b-1\over \b}{1\over z\ln
z}-{\a(z^{\a-1}-(z-1)^{\a-1})\over
z^{\a}-(z-1)^\a}\\
&=& {\b(\a-1) z\ln z+(\b-1)(z-1)\over \b
z(z-1)\ln z}-{\a(z^{\a-1}-(z-1)^{\a-1}\over
z^{\a}-(z-1)^\a}\\
&=&{\b z \ln
z\left[(z-1)^\a+(\a-z)z^{\a-1}\right]+(\b-1)(z-1)\left[z^\a-(z-1)^\a\right]
\over \b z (z-1) \ln z \left[z^\a-(z-1)^\a\right]}\\
&=& {s(z)\over  \b z (z-1) \ln z \left[z^\a-(z-1)^\a\right]}
\end{eqnarray*}

Hence the sign of $r'(z)$ is the same as the sign of $s(z)$ since $r(z)>0$ and $\b z (z-1) \ln z \left[z^\a-(z-1)^\a\right]>0$.

\begin{figure}
\centering\epsfig{file=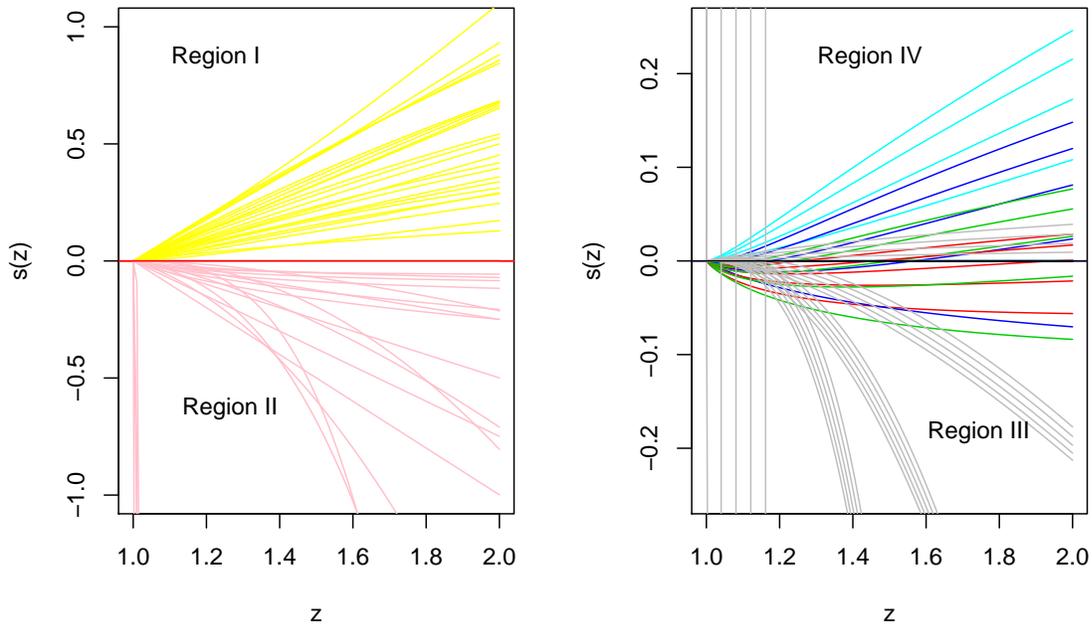, width=6in, height=4in}
\caption{The graphs of $s(z)$ for the four defined regions.
Region I=$\{(\alpha,\beta):
\beta\ge 1 \mbox{ and } \alpha\beta\ge 1\} $,  region II=$\{(\alpha,\beta): \beta\le 1 \mbox{ and
} \alpha\beta\le 1\} $, region
III=$\{(\alpha,\beta): \beta< 1 \mbox{ and } \alpha\beta< 1\} $ and region IV=$\{(\alpha,\beta): \beta> 1 \mbox{ and }
\alpha\beta< 1\} $, respectively.
}
\end{figure}

Figure 2 shows the graphs of $s(z)$ for parameters in the four regions.  From the left panel, one observes that $s(z)$ takes pure positive and pure negative values  in region I and region II, respectively, which implies that the hazard function is
monotone increasing in region I and monotone decreasing in region II. From the right panel, one observes that
$s(z)$ takes positive values first then drop to negative values in region III, which indicates the unimodal property
of the hazard function. In region IV, it is shown that $s(z)$ takes negative values first then positive values indicating
bath-tub shaped of the hazard function.  Furthermore, one notices that in region IV, the function
$s(z)$ may be positive or negative for all $z$ for some of the parameters $(\alpha, \beta)$. Overall, the shape of the hazard function is independent of the scale parameter.

\section{Maximum likelihood estimator under type II censoring}
Let $x_1,..., x_n$ be the lifetimes of the $n$ independent
components under testing. With type II censoring scheme,  one
observes the first $r$ order statistics, $x_{1:n}\le x_{2:n}\le
\cdots\le x_{r:n}$, of the sample $x_1,..., x_n$.
Based on the censored data $x_{1:n},..., x_{r:n}$, the likelihood function is
$$
L(\theta)={n!\over (n-r)!} \prod_{i=1}^r f(x_{i:n};
\theta)\left[1-F(x_{r:n}; \theta)\right]^{n-r},\ \  x_{1:n}\le
x_{2:n}\le\cdots\le x_{r:n}.
$$
The maximum likelihood estimator of $\theta$ satisfies the
following score equations:
\begin{eqnarray*}\label{score}
\pa ln L(\theta)&=&\sum_{i=1}^r {\pa f(x_{i:n};\theta)\over
f(x_{i:n};\theta)}-(n-r){\pa F(x_{r:n};\theta)\over
1-F(x_{r:n};\theta)}=0,
\end{eqnarray*}
where $\pa
f(x;\theta)=(f'_\a(x;\theta),f'_\b(x;\theta),f'_\s(x;\theta))^T$
and $\pa
F(x;\theta)=(F'_\a(x;\theta),F'_\b(x;\theta),F'_\s(x;\theta))^T$.
After some algebraic manipulations, we have
\begin{eqnarray}
{f'_\a(x;\theta)\over f(x;\theta)}&=&{1\over \a}\left[1+\ln F(x;
\theta)\right],\label{pa1}\\
 {F'_\a (x; \theta)\over
1-F(x;\theta)}&=&{1\over \a}{F(x;\theta)\ln F(x;\theta)\over
1-F(x; \theta)};\label{pa2}\\
%\end{eqnarray*}
%\begin{eqnarray*}
{f'_\b(x;\theta)\over f(x;\theta)}&=& {1\over
\b}\left\{1+\log\xs^\b
     \left[
      1-\xs^\b+{(\a-1)xf(x;\theta)\over \a\b F(x;\theta)}
     \right]
\right\},\label{pa3}\\
{F'_\b (x; \theta)\over 1-F(x;\theta)}&=&{1\over \b^2}\log\xs^\b
{xf(x;\theta)\over 1-F(x; \theta)}; \label{pa4}\\
%\end{eqnarray*}
%\begin{eqnarray*}
{f'_\s(x;\theta)\over f(x;\theta)}&=&-{\b\over
\s}\left[1-\xs^\b+{(\a-1)x f(x;\theta)\over \a\b
F(x;\theta)}\right],\label{pa5}\\
{F'_\s (x; \theta)\over 1-F(x;\theta)}&=&{1\over
\s}{xf(x;\theta)\over 1-F(x; \theta)}. \label{pa6}
\end{eqnarray}
Notice that if $\{x_i\}$ is a random sample from EWD with
parameter vector $\theta$, then $\{y_i=x_i^\b\}$ is a random
sample from exponentiated exponential distributed (EED) with shape
parameter $\a$ and scale parameter $\lambda=\s^\b$. Since $\b>0$, $x_{1:n}\le
x_{2:n}\le \cdots\le x_{r:n}$ is equivalent to $y_{1:n}\le
y_{2:n}\le \cdots\le y_{r:n}$, the first $r$ order statistics of
$y_1, \cdots, y_n$.

Hence we propose the following back fitting
algorithm to obtain the MLE of $\theta$:
\begin{itemize}
\item[Step 1.] For a given initial value $\b^{(0)}$ of $\b$, maximize $L(\a,\b^{(0)}, \s)$ with respect to $(\a,\s)^T\in R_{+}^2$. Denote the maximizer by $(\a^{(1)},\s^{(1)})^T$.
\item[Step 2.] Substitute $\a=\a^{(1)},\s=\s^{(1)}$ into $L(\a,\b,\s)$ to get profile likelihood function $L_1(\b)$. Maximize $L_1(\b)$ over $R_{+}$ to obtain $\b^{(1)}$.
\item[Step 3.] Repeat Steps 1 and 2 $k$ times to get $\theta^{(k)}=(\a^{(k)},\b^{(k)},\s^{(k)})^T$.
\item[Step 4.] Stop if $\|\theta^{(k)}-\theta^{(k-1)}\|<\epsilon$ for a pre-chosen small $\epsilon>0$.
\end{itemize}

\begin{theorem} The limit of the back fitting estimator is the
MLE. That is, \begin{equation}\label{thm1}
\max_{\b\in R_{+}}\max_{(\a,\s)^T\in R_{+}^2} L(\a,\b,\s)=\max_{\theta\in R_+^3} L(\theta).
\end{equation}
\end{theorem}

{\bf Proof:} Denote the MLE of $\theta$ by
$\hat{\theta}_{MLE}=(\hat{\a}_{MLE},\hat{\b}_{MLE},\hat{\s}_{MLE})^T$
and the estimator obtained by back fitting algorithm by
$\hat{\theta}_{BF}$.

It is clear that the right hand side of the  equation (\ref{thm1})
is bigger than or equal to the left hand side. For the other
direction, one notices by definition that
\begin{eqnarray*}
\max_{\theta\in R_+^3} L(\theta)&=&L(\hat{\theta}_{MLE})\le \max_{(\a,\s)^T\in R_{+}^2} L(\a,\hat{\b}_{MLE},\s)\\
&=&L(\hat{\a}_{BF},\hat{\b}_{MLE}, \hat{\s}_{BF})\le L(\hat{\a}_{BF},\hat{\b}_{BF}, \hat{\s}_{BF})\\
&=&\max_{\b\in R_{+}}\max_{(\a,\s)^T\in R_{+}^2} L(\a,\b,\s).
\end{eqnarray*}
This completes the proof.

%Then $\hat{\theta}_{MLE}=\hat{\theta}_{BF}$.

Hence for a given $\b>0$, the problem is equivalent to find the
MLE for EED with type II censored data $y_{1:n}=x_{1:n}^\b,\cdots,
y_{r:n}=x_{r:n}^\b$. In fact, the MLE of the EED parameters
satisfies the following fix point equation
\begin{equation}\label{iterative}
(\a,\lambda)=rg(\a,\lambda),
\end{equation}
 where
$g(\a,\lambda)=(g_1^{-1}(\a,\lambda),g_2(\a,\lambda)$ and
\begin{eqnarray*}
g_1(\a,\lambda)&=&(n-r){\left(1-e^{-y_{r:n}/\lambda}\right)^\a\ln (1-e^{-y_{r:n}/\lambda})\over 1-\left(1-e^{-y_{r:n}/\lambda}\right)^\a}-\sum_{i=1}^r\ln\left(1-e^{-y_{i:n}/\lambda}\right),\\
g_2(\a,\lambda)&=&\left[(n-r){\a y_{r:n} e^{-y_{r:n}/\lambda}\left(1-e^{-y_{r:n}/\lambda}\right)^{\a-1}\over 1-\left(1-e^{-y_{r:n}/\lambda}\right)^\a}
+(\a-1)\sum_{i=1}^r {y_{i:n}e^{-y_{i:n}/\lambda}\over 1-e^{-y_{i:n}/\lambda}}+\sum_{i=1}^r y_{i:n}\right]/r^2.
\end{eqnarray*}

To solve equation (\ref{iterative}), one can choose an initial value $\a^{(0)}$ and $\lambda^{(0)}$ for $\a$ and $\lambda$, respectively. Substitute $\a^{(0)}$ and $\lambda^{(0)}$ into the right hand side of the equation (\ref{iterative}) to get $\a^{(1)}$ and $\lambda^{(1)}$. Continue this procedure $k$ times to get   $\a^{(k)}=rg_1^{-1}(\a^{(k-1)}, \s^{(k-1)})$ and $\s^{(k)}=rg_2^{-1}(\a^{(k-1)}, \lambda^{(k-1)})$ for $k=0, 1,...$. The iteration stops when
$\|(\a^{(k)},\lambda^{(k)})-(\a^{(k-1)},\lambda^{(k-1)})\|<\epsilon$, a given pre-selected small positive number such as $10^{-8}$. For given $\b$, the estimator of $(\a,\lambda)$ is denoted by $(\hat{\a}, \hat{\lambda})$, hence the estimator of $\s$ is a function of $\b$, denoted by $\hat{\s}(\b)=\hat{\lambda}^{1/\b}$. Plug in the $\hat{\a}$ and $\hat{\s}(\b)$ into the log-likelihood function to obtain the profile likelihood function $L_1(\b)$ in $\b$. Maximizing $L_1(\b)$ to obtain $\hat{\b}$ and hence to obtain the maximum likelihood estimator $\hat{\theta}=(\hat{\a},\hat{\b}, \hat{\s})^T=(\hat{\a},\hat{\b}, \hat{\s}(\hat{\b}))^T$.

\section{Fisher information matrix}

Now we assume that $r/n\rightarrow p\in(0,1)$ as $n\rightarrow \infty$.
 Notice that the hazard function
$h(x;\theta)=f(x,\theta)/(1-F(x;\theta))$. Denote the Fisher
information matrix based on the first $r$ order statistics by
$I_p(\theta)$ and let $\lambda_p$ be the $100p$ percentile of
$F(x;\theta)$ such that $F(\lambda_p;\theta)=p$.

\begin{lemma} (Zheng, 2001)\label{l1}
Assume $F(x;\theta)$ has the same support for any $\theta\in
\Theta$, an open set in $R^k$. For $x\in R$ and $\theta\in
\Theta$, assume
$$
\pa f(x;\theta)={\partial^2\over \partial \theta \partial
x}F(x;\theta)={\partial^2\over \partial x \partial
\theta}F(x;\theta),
$$
where all derivatives exist. Furthermore, under the
interchangeability property for orders of limits, derivative and
integral,   the limiting Fisher information matrix can be
expressed as a single integral of hazard function under type II
censoring. That is,
$$
I_p(\theta)=\int_0^{\lambda_p}\Pa\Pa^T f(x;\theta)\ dx.
$$
Consequently, the asymptotic covariance matrix of the maximum
likelihood estimator of $\theta$ based on the first $r$ order
statistics is $I_p^{-1}(\theta)$.
\end{lemma}
Notice that
\begin{equation}\label{hazard}
\pah={\paf \over f(x;\theta)}+{\paF\over 1-F(x;\theta)}.
\end{equation}
Thus, for EWD family, the equations (\ref{pa1})-(\ref{pa6}) imply
that
\begin{eqnarray}
{\partial\over \partial\a}\ln h(x;\theta)
&=&{1\over\a}
\left[1+{\ln F(x;\theta)\over 1-F(x;\theta)}\right], \label{h1}\\
{\partial\over \partial\b}\ln h(x;\theta)
&=&{1\over \b}
\left\{
      1+\ln \xs^\b
  \left[
        1-\xs^\b+ {(\a-1)xf(x;\theta)\over \a\b F(x;\theta)}+{xf(x;\theta)\over
        \b (1-F(x;\theta))}
   \right]
\right\},\label{h2}\\
 {\partial\over \partial\s}\ln h(x;\theta)
 &=&
 -{\b\over\s}
 \left\{
      1-\xs^\b+{(\a-1)xf(x;\theta)\over \a\b F(x;\theta)}+{xf(x;\theta)\over
        \b (1-F(x;\theta))}
 \right\}.\label{h3}
\end{eqnarray}
Denote the Fisher information matrix
based on the first $r$ order statistics by
$$
I_p(\theta)=\left[I_p^{ij}(\theta)\right], i=1,2,3; j=1,2,3.
$$
Let
$$
\psi(z;\a)=1+\ln (1-z)\left[1+\left({1-z\over z}\right)\left(1-{\a\over 1-z^\a}\right)\right].
$$
Notice that $\psi(z;1)=1$.
 Then, we have our main theorem.

\begin{theorem}
Let $r/n\rightarrow p=F(\lambda_p;\theta)\in (0,1)$. For EWD
family with parameter vector $\theta=(\a,\b,\s)^T$ under type II
censoring, we have
\begin{eqnarray}
I_p^{11}(\theta)&=&{1\over\a^2} \int_0^p \left[1+{\ln x\over
1-x}\right]^2 \ dx,\label{i11}\\
I_p^{22}(\theta)&=&{\a\over \b^2}\int_0^{p^{1/\a}}
\bigg\{1+\ln[-\ln(1-x)]\psi(x;\a)\bigg\}^2 x^{\a-1}\ dx, \label{i22}\\
I^{33}(\theta)&=&\a\left({\b\over \s}\right)^2 \int_0^{p^{1/\a}}
\psi^2(x;\a) x^{\a-1}\
dx,\label{i33}\\
 I_p^{12}(\theta)&=&I_p^{21}(\theta)
 ={\a\over \b}\int_0^{p^{1/\a}}
 \left({1\over \a}+{\ln x\over 1-x^\a}\right)
 \bigg\{1+\ln [-\ln (1-x)]\psi(x;\a)\bigg\}
  x^{\a-1}\ dx,\\
I_p^{13}(\theta)&=&I_p^{31}(\theta)=-{\a\b\over \s} \int_0^{p^{1/\a}}\left({1\over\a}+{\ln x\over 1-x^\a}\right)\psi(x;\a)x^{\a-1}\ dx,\\
I_p^{23}(\theta)&=&I_p^{32}(\theta)=-{\a\over \s} \int_0^{p^{1\over \a}}\left\{
 1+\ln [-\ln (1-x)]\psi(x;\a)\right\}\psi(x;\a)x^{\a-1}\ dx.\label{i23}
\end{eqnarray}
\end{theorem}

{\bf Proof:} By Lemma \ref{l1}, equations (\ref{h1}) and $F(\lambda_p;\theta)=p$, we have
\begin{eqnarray*}
I_p^{11}(\theta)&=&\int_0^{\lambda_p} \left[{\partial\over \partial\a}\ln h(x;\theta)\right]^2\ dF(x;\theta)\\
&=&\int_0^{\lambda_p} {1\over\a}
\left[1+{\ln F(x;\theta)\over 1-F(x;\theta)}\right]\ dF(x;\theta) \\
&=&{1\over\a^2} \int_0^p \left[1+{\ln x\over
1-x}\right]^2 \ dx,
\end{eqnarray*}
which completes the proof of equation (\ref{i11}). To proof (\ref{i22}), we introduce new variable
by change of variables using $1-e^{-\xs^\b}=z$. Then
$F(x;\theta)=z^\a$, $ \xs^\b=-\ln (1-z)$,  $f(x;\theta)dx=\a z^{\a-1} dz$,
\begin{equation}\label{eq1}
{xf(x;\theta)\over \a\b F(x;\theta)}=-{(1-z)\ln(1-z)\over
z},
\end{equation}
\begin{equation}\label{eq2}
{xf(x;\theta)\over \b (1-F(x;\theta))}=-{\a z^{\a-1}(1-z)\ln(1-z)\over 1-z^\a},
\end{equation}
and
\begin{equation}\label{eq3}
1-\xs^\b+{(\a-1)xf(x;\theta)\over \a\b
F(x;\theta)}+{xf(x;\theta)\over
        \b (1-F(x;\theta))}=\psi(z;\a).
\end{equation}

Equations (\ref{h2}) and (\ref{eq3}) imply that
\begin{eqnarray*}
I_p^{22}(\theta)&=&\int_0^{\lambda_p} \left[{\partial\over \partial\b}\ln h(x;\theta)\right]^2\ dF(x;\theta)\\
&=&{\a\over \b^2}\int_0^{p^{1/\a}}
\left[1+\ln(-\ln(1-x))\psi(x;\a)\right]^2 x^{\a-1}\ dx,
\end{eqnarray*}
which completes the proof of (\ref{i22}).
Similarly, direct algebraic manipulations lead the other equations (\ref{i33})-(\ref{i23}).

Using integration and Taylor series expansion methods, one can verify that EWD satisfies the regularity conditions (Bhattacharyya, 1985). Hence we have the asymptotic normality theorem.

\begin{theorem}
Let $r/n\rightarrow p=F(\lambda_p;\theta)\in (0,1)$. For EWD
family with parameter vector $\theta=(\a,\b,\s)^T\in \Theta$ under
type II censoring, we have
$$
\sqrt{n}\left(\hat{\theta}_{MLE}-\theta\right)\rightarrow N_3(0,
I_p^{-1}(\theta)), \mbox{ as $n\rightarrow\infty$}.
$$
\end{theorem}

\section{Real data analysis: two examples}

In this section, we use maximum likelihood method to fit two real data sets. One is the ball bearings lifetime analyzed in Gupta and Kundu (2001) and the other is the breaking stress of carbon fibres (in Gba) from Nichols and Padgett (2006). For the first data set, Caroni (2002) has pointed out that the data set contains censored points. For the second data, it will be interesting to see how the censoring rate affect the estimation. We consider three censoring rates of no censoring, 10\% censoring and 20\% censoring. The ball bearings lifetime data set contains 23 observations, while the breaking stress of carbon fibres data contains 100 observations. We fit the data sets using both the exponentiated exponential distribution (EED) and  the exponentiated Weibull distribution (EWD). For the maximum likelihood estimates of EED, a modified quasi-Newton method with box constraints in the function {\it optim()} in {\bf R} package is used. The maximum likelihood estimates are presented in Table 1. Note that $EED(\a,\s)=EWD(\a,1,\s)$.

\begin{center}
Table 1: Maximum likelihood estimates of the model parameters\\
 for ball bearings lifetime\\

\begin{tabular} {|l|l|r|r|r|} \hline
             &          & \multicolumn{3}{|c|}{Censoring rate}\\ \hline
Distribution & Estimate &  $0\%$  & $10\%$  & $20\%$\\  \hline
\multirow{2}{*}{EED}          & $\alpha$ &  5.2707 & 5.0752  &  5.0728\\
             & $\sigma$ & 31.0035 & 31.7540 &  31.7592\\
             & -log(likelihood)& 112.9762& 104.6143& 91.0536\\ \hline
\multirow{3}{*}{EWD}          & $\alpha$ & 4.7446  & 7.7412  &  9.0634\\
             & $\beta$  & 1.0444  & 0.8462  &  0.7924\\
             & $\sigma$ & 33.6008 & 22.3618 & 19.3368 \\
             & -log(likelihood)  & 112.9740& 104.5917&  91.0128\\    \hline
\end{tabular}
\end{center}

\begin{figure}\label{contour}
\centering\epsfig{file=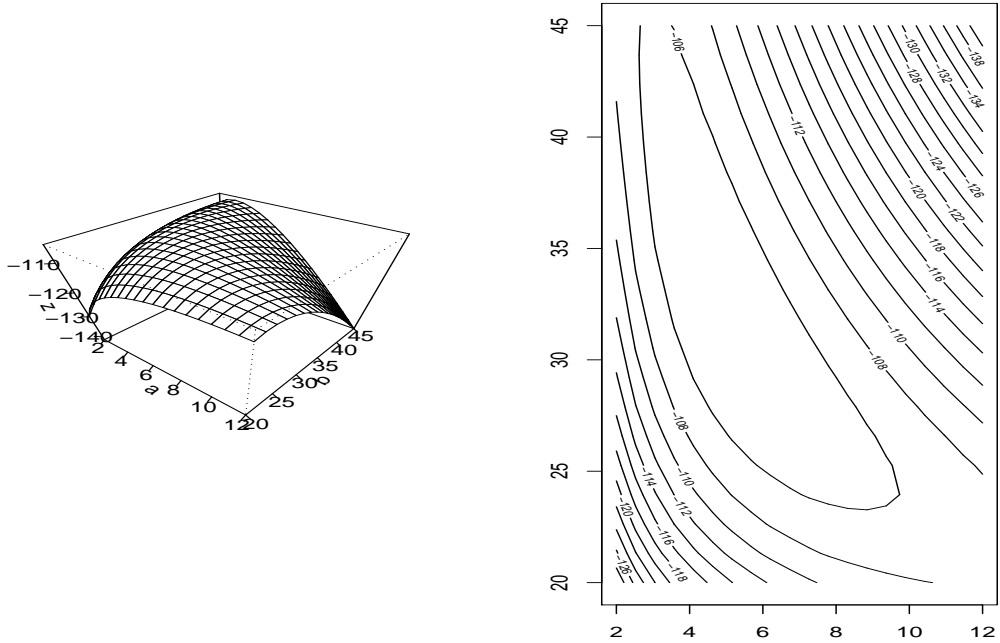, width=6in, height=4in}
\caption{The log-likelihood function and its contour plot of EED under 10\% censoring rate for ball bearings lifetime data.}
\end{figure}

The maximum likelihood estimator is robust against various censoring rates  for EED. On the other hand it is  sensitive for EWD. The standard likelihood ratio test shows that  the shape parameter $\b$ in EWD is not significant different from one, hence the EED is suitable in modeling the ball bearings lifetime. The result is consistent with Gupta and Kundu (2001) for no censoring data. This property still holds under various censoring data. As an illustration, with 10\% censoring rate, the log-likelihood function and its contour plot is given in Figure \ref{contour}.\\

\begin{center}
Table 2: Maximum likelihood estimates of the model parameters \\
for break stress data\\
\begin{tabular} {|l|l|r|r|r|} \hline
             &          & \multicolumn{3}{|c|}{Censoring rate}\\ \hline
Distribution & Estimate &  $0\%$  & $10\%$  & $20\%$\\  \hline
\multirow{2}{*}{EED}          & $\alpha$ &  7.7883 & 7.6053  & 6.9949\\
             & $\sigma$ & 0.9870 & .9994 & 1.0487\\
             & -log(likelihood)& 146.1823& 137.4110& 130.8363\\ \hline
\multirow{3}{*}{EWD}          & $\alpha$ & 1.3169  & .4432  & .1840\\
             & $\beta$  & 2.4091  & 5.5320  & 12.4404\\
             & $\sigma$ & 2.6824 & 3.4164 & 3.6032 \\
             & -log(likelihood)&  141.3320& 130.5830& 125.6935\\ \hline
\end{tabular}
\end{center}

\begin{figure}\label{contour1}
\centering\epsfig{file=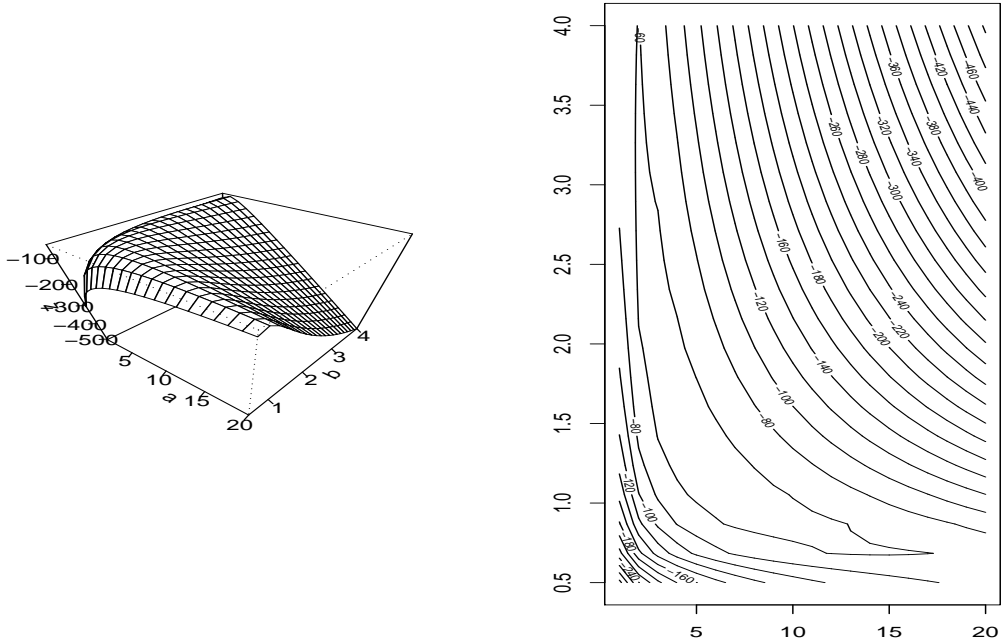, width=6in, height=4in}
\caption{The log-likelihood function and its contour plot of EED under 10\% censoring rate for the break stress data.}
\end{figure}

Once again, the maximum likelihood estimator is robust against various censoring rates for EED. On the other hand it is  sensitive for EWD. The shape parameter $\b$ in EWD is significant different from one by the standard likelihood ratio test. Hence it is better to use EWD in modeling the break stress. The result is consistent with Nichols and Padgett (2006). The EWD fit still holds for censored data.  As an illustration, with 10\% censoring rate, the log-likelihood function and its contour plot is given in Figure \ref{contour1}.

\section{Conclusion}
For data generated from exponentiated Weibull distribution, we visualize the shape of the hazard function under various shape and scale parameters. Under
type II censoring, we propose a simple algorithm for computing the maximum likelihood estimator and derive the Fisher information matrix. The latter one is represented through a single integral in terms of hazard
function, hence it solves the problem of computation difficulty in
constructing inference for the maximum likelihood estimator. Data analysis for two real data sets shows that the maximum likelihood estimator is robust with respect to the censoring when the underlying distribution is the exponentiated exponential, but not for general exponentiated Weibull distribution when the shape parameter $\b\neq 1$.

\end{document}